# Understanding, Quantifying, and Controlling the Molecular Ordering of Semi-conducting Polymers: From Novices to Experts and Amorphous to Perfect Crystals

Zhengxing Peng, Long Ye[§,*], Harald Ade*

Department of Physics and Organic and Carbon Electronics Laboratories (ORaCEL), North Carolina State University, Raleigh, North Carolina 27695, United States

* E-mail: hwade@ncsu.edu

[§]Current address: School of Materials Science & Engineering and Tianjin Key Laboratory of Molecular Optoelectronic Science, Tianjin University, Tianjin 300350, P.R. China

* E-mail: yelong@tju.edu.cn

**Abstract:** Molecular packing, crystallinity, and texture of semiconducting polymers are often critical to performance. Although frame-works exist to quantify the ordering, interpretations are often just qualitative, resulting in imprecise and liberal use of terminology. Here, we reemphasize the continuity of the degree of molecular ordering and advocate that a more nuanced and consistent terminology is used with regards to crystallinity, semicyrstallinity, paracrystallinity, crystallite/aggregate, and related characteristics. We are motivated in part by our own imprecise and inconsistent use of terminology and the need to have a primer or tutorial reference to teach new group members. We show that a deeper understanding can be achieved by combining grazing-incidence wide-angle X-ray scattering and differential scanning calorimetry. We classify a broad range of representative polymers into four proposed categories based on the quantitative analysis of molecular order based on the paracrystalline disorder parameter ($g$). A small database is presented for over 10 representative conjugated and insulating polymers ranging from amorphous to semicrystalline. Finally, we outline the challenges to rationally design perfect polymer crystals and propose a new molecular design approach that envisions conceptual molecular grafting that is akin to strained and unstrained hetero-epitaxy in classic (compound) semiconductors thin film growth.

## 1. Introduction

In the past two decades, thousands of new π-conjugated materials have been reported and used in organic electronic devices. Polymer semiconductors offer great opportunities and potentials due to their light-weight, color tunability, mechanic flexibility and the ability to be deposited on large-area flexible substrates at low cost, and providing active materials for a range of electronic applications, such as organic field-effect transistors[1], organic electrochemical transistors[2], organic light-emitting diodes[3] and organic solar cells[4]. The intrinsic properties of polymer chains, such as different degrees of conformational and configurational freedom, determine how crystalline or amorphous the polymeric materials are[5, 6]. The degree of crystallinity and perfection of the ordering is determined by the local molecular packing, and the resulting ordered or disordered films and their texture affect optical absorption, exciton delocalization, charge separation, and charge transfer. In organic electronics, certain high-performing materials often tend to pack in an ordered molecular arrangement, such as crystallites or aggregates.

Conjugated polymers are generally homopolymers or copolymers (see **Figure 1a**) and often identified and reported in a somewhat vague manner to be amorphous or semicrystalline. Sometimes, the terminology is not well defined and not consistent across the literature. For example, the benchmark and widely studied polymer PTB7 (full name see appendix A) of interest in solar cells was reported to be semicrystalline in several papers[7, 8] because of the presence of (100) and (010) reflection peaks in the X-ray diffraction patterns, while in other papers PTB7 was considered amorphous[9, 10] due to the lack of the long-range order. Whatever the cause of this inconsistency and given existing basic IUPAC definitions[11] and prior work utilizing the Warren-Averbach framework[12, 13], it is highly desirable that a more consistent terminology be used even when only semi-quantitative analysis is performed.

To help guide and encourage more precise use of terminology, we discuss the molecular ordering of conjugated polymers and identify the polymers to be amorphous or semicrystalline in a quantitative manner by combining both grazing-incidence wide-angle X-ray scattering (GIWAXS)[14] and differential scanning calorimetry (DSC). We start by discussing important concepts such as crystals, crystallites and crystallinity in polymeric materials, followed by the discussion on GIWAXS and DSC of the nominally amorphous polymers, atactic polystyrene (PS) and poly (methyl methacrylate) (PMMA), and the nominally semicrystalline polymer, poly(3-

hexylthiophene) (P3HT). This clarifies characteristics of semicrystalline and amorphous polymers. We advocate use of a nomenclature that differentiates into 3D amorphous, 2D amorphous, aggregates (short-range order) and crystallites (long-range order) based on the paracrystalline disorder parameter (*g* parameter) and thermal characteristics.

We furthermore review recent work that clearly delineates that sidechain ordering and backbone ordering is generally not synergistic, a competition that is likely at the very core of semiconducting polymers exhibiting generally large *g* parameters of >8%. Questions naturally arise as to what it would take to achieve nearly perfect crystals of semiconducting polymers with disorder that approaches that of TIPS-pentacene ($g < 2\%$). Solutions of highly ordered materials with long-range ordering and low *g* parameter would consequently require special molecular design that is possibly guided by the novel conceptual hetero-epitaxy grafting design method proposed here.

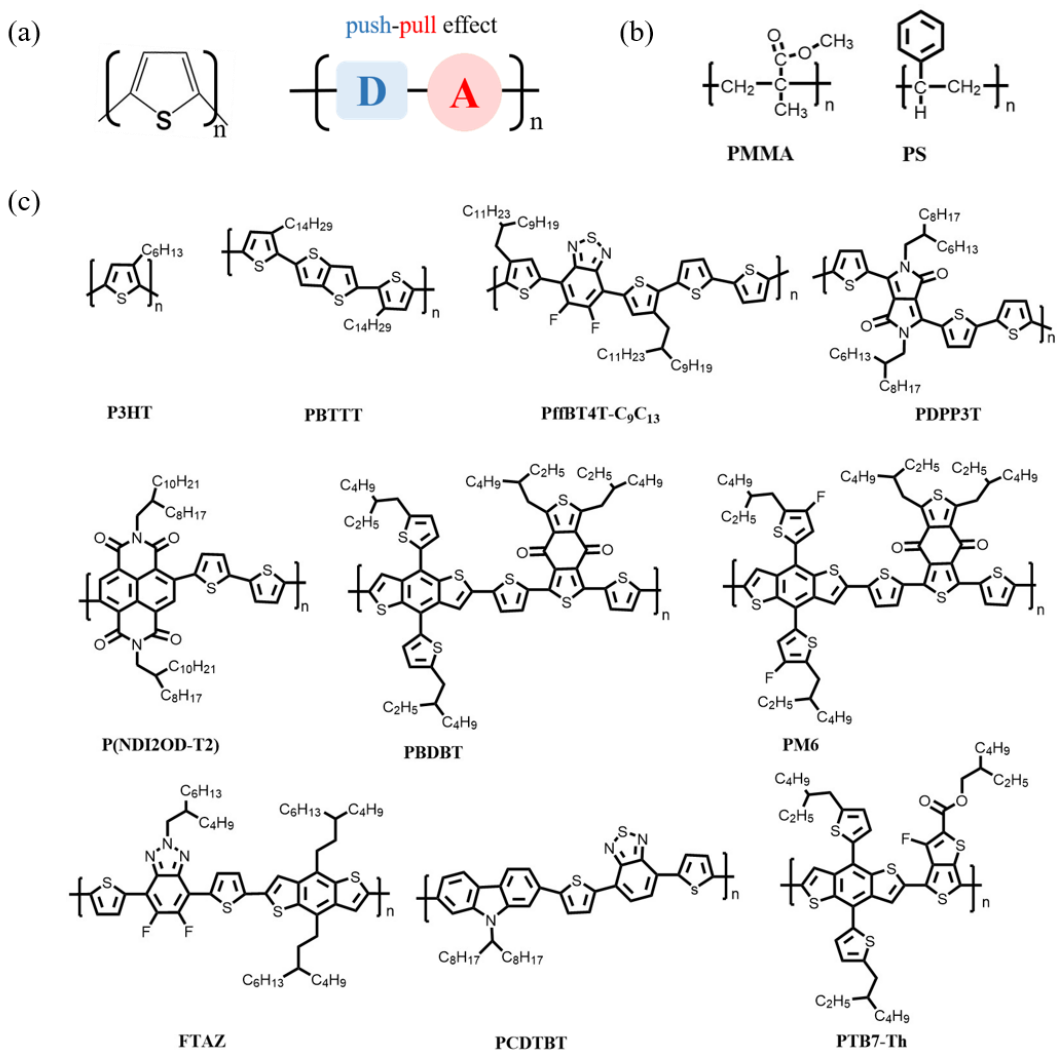

**Figure 1.** (a) The two types of conjugate polymers (left: polythiophenes, which may have various sidechains, right: donor–acceptor (D–A) alternating conjugated copolymers with push-pull effect). (b) The chemical structure of two amorphous nonconjugate reference polymers. (c) The chemical structure of the conjugate polymers used here.

## 2. Quantification of Molecular Order and Related Terminologies

### 2.1 Clarifications of the key concepts related to molecular order

In solid state matters, a crystal is defined to be a periodic array of identical motifs, suggesting a long-range positional order of the atomic length scale in three-dimensional space. However, in polymeric materials, there might be ordering that lacks periodicities in at least one dimensions[5], as three-dimensional long-range order is almost impossible. Also, there always exists structural disorder, paracrystallinity or defects inside polymeric crystals, and thus, polymeric materials are never completely crystalline. Due to kinetics and the covalent constraint along the backbone, polymers often have large amorphous volume fractions in bulk materials and even in the most highly ordered materials are referred to as semicrystalline. Since the ordered phase in a typical semicrystalline polymer film is on the order of 1-100 nm in diameter or thickness[12], such ordered regions are usually called crystallite rather than crystal. Aggregate is also used frequently while its definition is vague and needs to be clarified. In a broad sense, 'aggregate' refers to clusters of the material owing to strong intermolecular or intramolecular interaction between the molecules. In the spectral aspect, the aggregates in solution or thin films are directly associated with changes in the emission/absorption spectral line shape[15], as compared to the unaggregated states. Typically, optical aggregate can be classified to be J-aggregate and H-aggregate, where the J-aggregate refers to the appearance of an intense, red-shifted narrow band in the absorption spectrum and the H-aggregate refers to the main absorption peak to be blue-shifted[16]. In the X-ray diffraction view, D. T. Duong *et al.*[17, 18] defined an "aggregate" as a group of the π-stacked conjugated segments with the lack of the lamellar stacking and thus a crystallite contains aggregates but not all aggregates are crystallites. With enough π-stacked conjugated segments, aggregates can also produce discernible diffraction peaks. R. Noriega *et al.*[19] also proposed that small semi-ordered domains with short-range ordering of a few molecular units to be referred to as aggregates, compared to the crystallites with larger domains and with better three-dimensional long-range periodicity. Short-range ordered aggregates often yield sufficient π-orbital overlap[15, 20] to facilitate interchain charge transport, resulting in some seemingly contradictory cases where the polymers show a less ordered lamellar packing structure but excellent charge transport properties[19]. This contradiction indicates that the ordering along the lamella and π- π stacking is not always synergistic. Not only is the range of the ordering important, but also the direction in which the ordering is most pronounced. Diffraction and calorimetry generally probe crystallites while optical measurements tend to probe aggregates[18].

A typical microstructure of a semicrystalline polymer film is composed of crystallites (lamellae) embedded in an amorphous phase, producing a highly interconnected network. Since there exist both ordered and disordered or amorphous regions, the fraction of crystalline to amorphous regions is a critical parameter quantifying how crystalline a polymer is. The crystallinity or the degree of crystallinity (DoC) is defined as the volume fraction of crystalline material in a film[12]. In X-ray diffraction experiments, the integrated intensity of a diffraction peak is often proportional to the amount of crystalline material in a thin film, and thus can quantify the DoC. However, an absolute DoC is close to impossible to be quantified since it would require reference samples to be entirely crystalline and entirely amorphous films, which is hardly obtained in polymeric materials. Instead, the relative DoC[12, 21, 22], describing the DoC of one film as compared to another film of the same material, can be easily quantified.  Generally, the relative DoC (rDoC) can be quantified by several methods, such as GIWAXS, DSC, nuclear magnetic resonance (NMR) and density via dilatometer. To characterize the DoC via GIWAXS, J. Balko *et al.*[23] proposed that the DoC can be estimated by unity minus the ratio of the scattering intensity at a certain scattering vector (which indicates the percentage of amorphous components), as shown in **Figure 2a**, in between Bragg reflections of the semicrystalline sample to that in a completely amorphous (molten) sample. Another way of using x-ray diffraction is to integrate the intensity of a Ewald sphere corrected pole figure from out-of-plane direction to in-plane direction, which is proportional to a film's degree of crystallinity[24], and then to compare between different samples. Also, the rDoC of two specimens of the same material can also be characterized by the ratio of their specific enthalpy of fusion via DSC[23]. The possibility for NMR to determine DoC arises through an intrinsic difference between the nuclear resonance frequency of the targeted nuclei, such as $^1$H and $^{13}$C, of the ordered domains and the amorphous domains. After decomposition of the resonance spectra into the two parts of ordered and amorphous components, the DoC is given by the percentage of the intensity of the ordered components[21, 25]. Yet another way to characterize the rDoC is the dilatometric methods, which can be applied to measure the specific volume of the materials. The resulted specific volume is assumed to be the sum of the percentage of the specific volume of the crystalline and amorphous components while the specific volume of the amorphous components can be estimated from the molten state[26]. These methods do not precisely measure though the same parameter, as any density measurement, for example, is also impacted by the relative ratio of the liquid and rigid amorphous phase[27].

In the literature, using the term "crystallinity" without qualifiers is often confusing because two different concepts, degree of crystallinity and quality of ordering, within a crystallite are mixed up. Sometimes, crystallinity is used to describe the quality of crystalline or semicrystalline ordering of organic materials, rather than referring to the degree of crystallinity. For example, when analyzing and discussing the GIWAXS data, it was claimed that the two scattering peaks, namely the so-called lamellar (100) stacking peak and (010) π-π stacking peak, of PBDT-TDZ and PBDTS-TDZ films showed their crystallinity nature[28]. In such cases, we advocate to use "crystalline" instead of "crystallinity" to preserve use of crystallinity referring to the volume fraction of crystalline to amorphous regions in a film. Quite often, crystallinity is used to refer to the quality of ordering of molecular packing or even just to spacing. For example, it was reported[28] that the shorter π-stacking distance of PBDTS-TDZ than PBDT-TDZ indicates a better crystallinity of PBDTS-TDZ. Or the observed (001) peak indicated the excellent crystallinity of P(NDI2OD-T2) when processed with 2-methyltetrahydrofuran[29]. Such use is prevalent and we do not intent to single anybody out, but simply provide examples and motivate our suggestion to use a reference to "molecular packing" or "molecular ordering" instead of "crystallinity", particularly in cases where only aggregates with short-range ordering are observed.

Some crystalline or semicrystalline materials have the ability to form more than one packing motifs in the solid or aggregated state, which are referred to as polymorphs[30]. The study of polymorphism attracts a lot of interest from the community since polymorphs can have different physical and thermodynamic properties, such as optical absorption, emission, electrical conductivity, solubility, and thermal stability. The polymorphism of conjugated polymers has been reported for various polymers. For example, M. Brinkmann *et al.*[31] observed two forms of polymorph for the P(NDI2OD-T2). In form I, adjacent chains show a segregated structure with donor (acceptor) units stacking on top of each other, while form II has a mixed structure with donor units stacking on top of the acceptor units of the adjacent chain. F. Peter *et al.*[32] have reported that regioregular oligo(3-hexylthiophene)s $(3HT)_n$ can feature two distinctly different solid-state structures, the polymorph form I and form II, with dissimilar crystal structures and thus different equilibrium melting temperatures as well as varied optical and electronic properties. M. Li *et al.*[33] reported that a low-bandgap diketopyrrolopyrrole polymer can form two distinctly different semicrystalline polymorphs $β_1$ and $β_2$ by controlling the solvent quality and the transition between the two polymorphisms via the amorphous α phase is observed.

A completely amorphous film morphology shows no ordering. However, it is hard to make a completely amorphous film with polymeric materials or without any molecular correlations between nearest molecular neighbors. The classical amorphous polymers, such as atactic polystyrene (PS) and poly (methyl methacrylate) (PMMA), tend to scatter diffusely in X-ray diffraction experiments with broad scattering peaks and will be discussed in detail below. For the conjugated polymers, the polymers that are disordered in other directions except for some short-range order in the π-stacking direction and with no melting peak in DSC scan, such as FTAZ[34], can also be identified to be largely amorphous. The central discussions below draw heavily on well understood x-ray scattering concepts that have been previously delineated by the Salleo group[12, 13, 19], which in turn harked back to earlier. For example, R. Noriega *et al.*[19] proposed to classify polymers with broad, featureless scattering peaks in X-ray diffraction experiments to be 3D amorphous materials. We aim to translate prior understanding to the organic device community at large by contextualizing and exemplifying important diffraction concepts broadly by comparing semiconducting polymers with various degree of ordering to the classic amorphous PS and PMMA and by comparing and pairing WAXS with DSC.

The ordering probed by GIWAXS and DSC is on the nanoscale with fixed spatial relations in various directions. However, there is also liquid crystalline ordering on the mesoscale, which is also important when one considers molecular ordering in semiconducting polymers. Liquid crystals are materials that have the properties of both a crystal and a liquid, which lies in an intermediate state between the amorphous state and crystalline state. The liquid crystals do not show the 3D positional order as the crystals do while retain some orientational order. For conjugated polymers, there are some polymers that show the liquid crystalline properties. X. Zhang *et al.*[35] have demonstrated that the resulting orientation map via dark-field transmission electron microscopy (DF-TEM) implies the in-plane PBTTT crystallite orientation varying smoothly across a length scale and significant in a relatively long range while only small angle variations between adjacent diffracting regions, exhibiting an in-plane liquid crystalline texture. The liquid crystalline texture helps to decrease the density of abrupt grain boundaries, leading to a relative insensitivity in organic thin-film transistor device properties. Soft x-ray scattering with polarized light (P-SoXS) can assess with high orientational sensitivity mesoscale ordering in the 20-2,000 nm range and has observed such characteristics even in as-case PBTTT film where TEM analysis was inconclusive[36, 37]. Length scales of >30 nm can be also readily probed with dichroic scanning transmission x-ray microscopy[38, 39]. R. Xie *et al.*[40] reported the nematic ordering for PFTBT and PCDTBT by a combination of DSC, temperature-dependent x-ray scattering and linear viscoelastic rheology. The local chain alignment via the nematic order can reduce the chain entanglement, leading to faster chain relaxation from the topological constraints of surrounding chains. We will not focus on liquid crystalline ordering here and mention only briefly these properly to provide completeness.

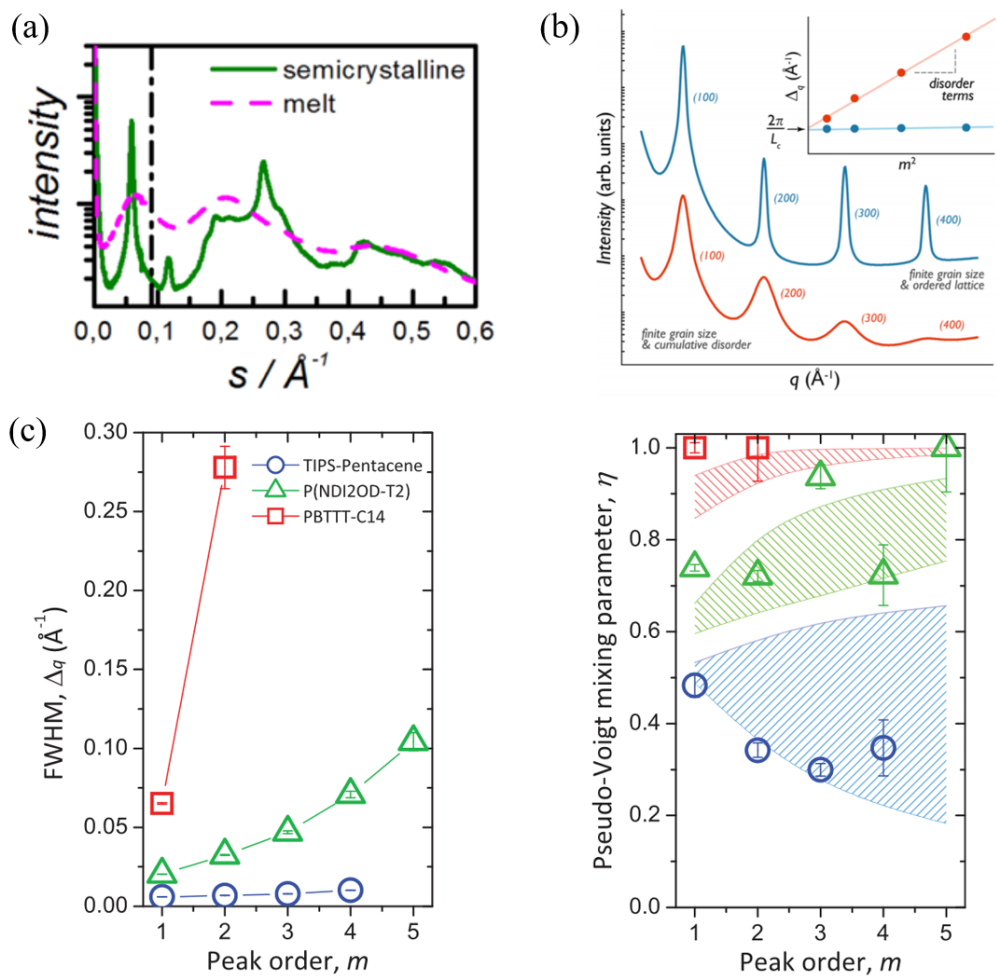

**Figure 2.** (a) SAXS/WAXS patterns for P3HT in the semicrystalline state at 40 °C (solid line) after cooling from the melt and in the melt (dashed line). The vertical dot-dashed line indicates the scattering vector to calculate the percentage of amorphous components with the ratio of the intensity in the semicrystalline sample to that in the melt sample, and thus DoC can be calculated accordingly. Reproduced with permission from ref. 23. (b) Schematic of hypothetical diffraction profile from crystalline lattices affected by finite size with (red) and without (blue) the contribution of cumulative disorder. Inset: The contribution of cumulative disorder results in an increasing peak width (red) while the absence of it gives a constant peak width (blue). Reproduced with permission from ref. 12. (c) Peak parameters determined from isolated x-ray diffraction peaks as a function of peak order for TIPS-pentacene (blue circle), P(NDI2OD-T2) (green triangle), and PBTTT (red square). Reproduced with permission from ref. 13.

## 2.2 Quantitative Characterization of Molecular Order

To quantitatively characterize the molecular order, GIWAXS or DSC are usually employed. GIWAXS[41-43] is the most common X-ray techniques used with organic semiconductors, probing the molecular packing and ordering in a thin film on the length scale of angstrom to tens of nanometers as well as the film texture. DSC is a powerful tool in detecting the crystalline phase of the film by the presence of melting/crystallization peaks. The parameters related to the molecular order characterized from GIWAXS and DSC include stacking distance ($d$), coherence length ($L_c$), number of stacking layers ($n$), crystallite size, paracrystalline disorder parameter ($g$), melting enthalpy ($\Delta H$) and relative DoC. We will review some of the basic aspects of these parameters below.

In X-ray scattering experiments, the deviation of the diffraction peaks from the ideal delta-function distribution of infinite, perfect crystals originates from the general disorder and finite size of the ordered domains, whose respective contributions can be disentangled by the diffraction line shape and intensity analysis if multiple peaks are present (see **Figure 2b** and **2c**). The lattice disorder can be classified in two forms: i) the noncumulative disorder (random statistical fluctuations about an ideal lattice position), which is characterized by the lowering of the peak intensity with no effect on the peak width/breadth (e.g. Debye–Waller factor for thermal fluctuations), and ii) the cumulative disorder (dislocations, impurities, chain backbone twists, or nonideal packing) that impacts the breadth and shape of the diffraction peaks, with the contribution of paracrystalline disorder (quantified by $g$ parameter, a measure of the percentage of statistical deviation from the mean lattice spacing in a crystal) and lattice-parameter fluctuations ($e_{rms}$)[13]. The line shape analysis involving the entire peak shape in the form of a Fourier-transform of the diffraction peaks was introduced by Warren and Averbach[44, 45]. The power of Warren-Averbach approach relies on the fact that the Fourier coefficients of the diffraction peaks can be decoupled into the product of contributions from a finite crystallite size, which is independent of peak order, and disorder terms that are peak-order dependent. By fitting the Fourier transforms of the isolated diffraction peaks with appropriate background subtraction, the Fourier coefficient gives the size- and disorder-dependent terms. Since the Warren-Averbach approach requires in-depth data processing and analysis, some simpler methods to estimate the size- and disorder-dependent terms are usually applied in practice.

A commonly used method to determine the crystallite size or coherence length is use of the Scherrer equation[46] via $L_c = \frac{2\pi K}{\Delta_q}$ where $\Delta_q$ is the full width at half maximum (FWHM) and K is the shape factor with a typical value of ~0.9. When the broadening of the diffraction peaks is mainly from finite crystallite size with weak disorder, the Scherrer equation is a good approximation for the crystallite size. However, in most cases of conjugated polymer systems, strong disorder dominates the peak broadening and thus the concept of the crystallite size is not valid anymore. Although the Scherrer equation works with the assumption that the crystallite size is the main contributor to the broadening of diffraction peaks while lattice disorder is ignored, it can still give descriptive evidence for relative changes when processing conditions are varied experimentally[13]. To distinguish the contribution of finite size from the cumulative disorder to the broadening of peaks, one practical way of peak shape analysis is to plot the peak width as a function of peak order. The schematic cases shown in **Figure 2b** illustrate that when the finite size is the only contributor to the broadening of peaks, the peak width is a constant as a function of peak order. If there is also a contribution from cumulative disorder, the peak width shows an increasing trend with peak order. An experimental example is shown in **Figure 2c** where TIPS-pentacene is dominated by size effect with a roughly constant peak width while cumulative disorder is observed in P(NDI2OD-T2) and PBTTT with increasing peak width with peak order. The peak shape analysis can also help to differentiate the paracrystalline disorder ($g$) from lattice-parameter fluctuation ($e_{rms}$) by exploring the pseudo-Voigt mixing parameter $\eta$ since the paracrystalline term contributes to a Lorentzian distribution while the lattice-parameter fluctuation is a Gaussian shape with the assumption that the lattice disorder is Gaussian random[13]. The fraction that comes from the Lorentzian function is given by a mixing parameter $\eta$. $\eta$ close to 1 (Lorentzian) indicates $g$-dominated disorder while $\eta$ close to 0 (Gaussian) suggests $e_{rms}$-dominated disorder. In the cases where $\eta$ is not near 0 or 1, both paracrystallinity and lattice-parameter fluctuations exist. An example is shown in **Figure 2c.** TIPS-pentacene has an $\eta<0.5$ and a decreasing trend, suggesting its highly crystalline behavior (very low $g$). P(NDI2OD-T2) shows a relatively constant $\eta\sim0.75$, illustrating a competition between paracrystallinity and lattice-spacing fluctuations. PBTTT with $\eta \sim 1$ shows strong paracrystallinity.

In practice, the paracrystalline disorder parameter $g$ can be estimated from the width of a single-peak for highly disordered system where the effects from lattice-parameter fluctuation $e_{rms}$ can be neglected via $g \approx \frac{1}{2\pi}\sqrt{\Delta_q d_{hkl}}$, where $\Delta_q$ and $d_{hkl}$ are the FWHM and the interplanar spacing of the diffraction peak of interest. When $e_{rms}$ can be neglected, this procedure is a good estimate for $g$ parameter. For example, PBTTT shows strong paracrystallinity with $\eta \sim 1$, and thus the $g$ parameter of (0*k*0) from WA analysis (~7.3%)[13] and from the single-peak analysis (8%) is close. However, when effects from lattice-parameter fluctuation are comparable to or even greater than paracrystallinity, the single-peak analysis is not applicable. For example, the *g* parameter for P(NDI2OD-T2) (*h*00) peaks and TIPS-pentacene (00*l*) peaks from the single-peak analysis are 14%, 5% respectively, which are not consistent with that from WA analysis (4%, 0.3%)[13]. For the polymers discussed in the next part, the effects from $e_{rms}$ in (010) peaks can mostly be neglected since their $\eta$ is close to 1, such as PTQ10 (~1), PBDB-T (~1), P(NDI2OD-T2) (~1), PffBT4T-2OD (~0.96), PTB7-Th (~0.94) and DPP3T (~0.80), and thus the single-peak analysis is applicable. In contrast, the $e_{rms}$ effects in (*h*00) cannot be neglected in most cases such as DPP3T, PM6, PTQ10 (as shown in **Figure 4n**), and thus the single-peak analysis cannot be performed.

Generally, *g* = 0% indicates a perfect crystal and 0% < *g* < 2% represents crystalline ordering while 2% < *g* < 12% is paracrystalline ordering; Amorphous silicon dioxide glass has *g* ≈ 12% and thus *g* > 12% is referred to the amorphous ordering[13, 47]. The stacking distance, or layer spacing, *d* of a certain diffraction peak, such as lamellar stacking peak or π-π stacking peak, can be calculated from the reciprocal of the peak position with $d = \frac{2\pi}{q}$. This stacking distance indicates how close the molecules pack together. The number of stacking layers (*n*) is characterized by the ratio of the coherence length ($L_c$) and the stacking distance (*d*) of a certain diffraction peak.

## 3. Classification of Polymers

Combining GIWAXS and DSC, we classify the polymers as described below into four categories, based on the degree of disorder: 3D amorphous, oriented ("2D") amorphous, semicrystalline with short-range order (aka aggregates), and semicrystalline with long-range order (aka crystallites). The overall classification scheme is illustrated and displayed in **Figure 3**. One key dividing line is the presence or absence of a melting transition in DSC and the presence of higher order diffraction peaks in GIWAXS. We will discuss these classes and their defining characteristic in turns using a number of representative materials.

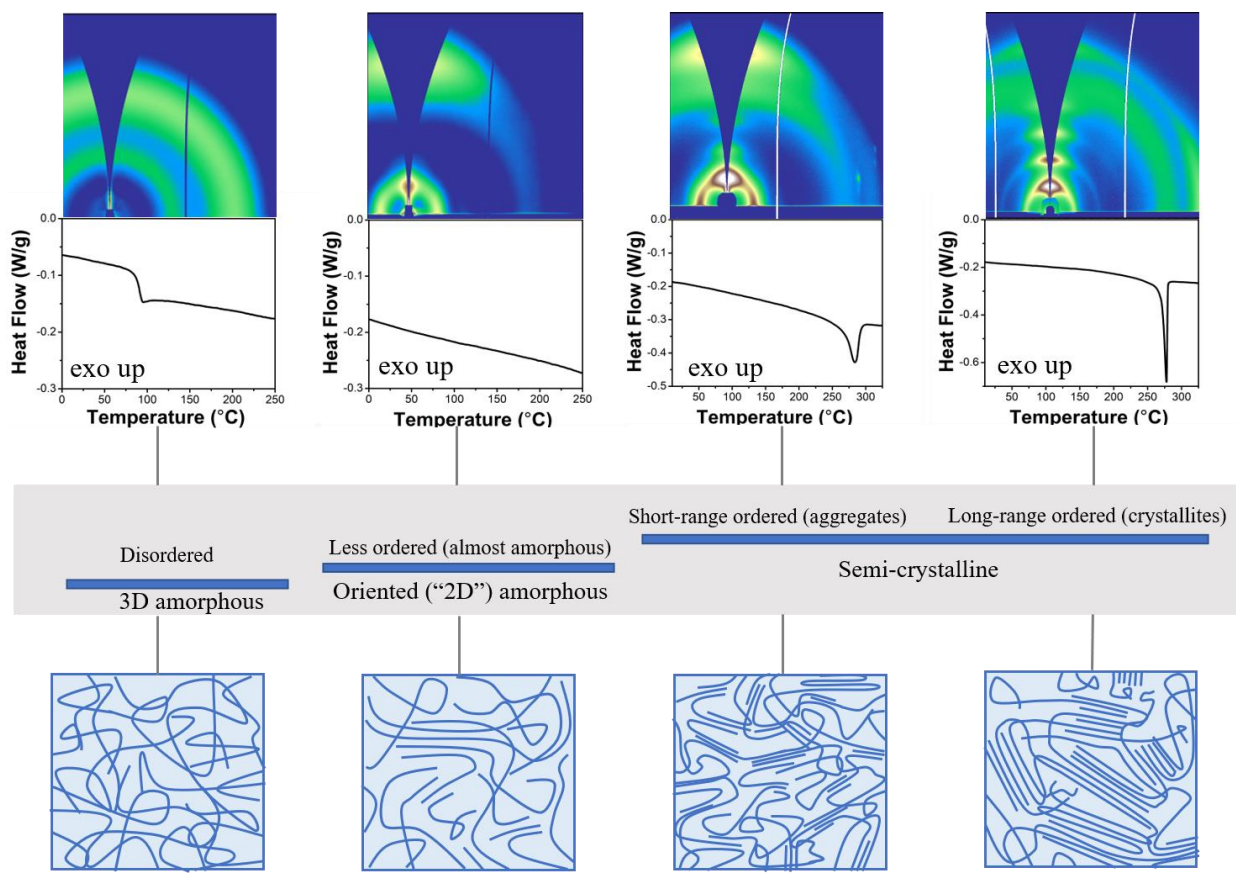

**Figure 3.** Qualitative classification of molecular ordering of a polymer material on the quantitative order-disorder scale using both GIWAXS and DSC criteria and schematic representations for various ordering. The 3D amorphous materials show no melting peak in DSC and broad, featureless scattering peaks in the GIWAXS pattern. The preferentially oriented ("2D") amorphous conjugated polymers only show lowest order, broad peaks in the GIWAXS pattern (indicating local short-range order only) with no melting peak in DSC heating scan. The semicrystalline material can be characterized by a melting peak observed in DSC and either short-range or long-range ordered are characterized according to the GIWAXS data and *g* parameter analysis.

For 3D amorphous polymers, we start the discussion with a pair of classical, amorphous polymers, atactic PS and PMMA, and then extend to representative amorphous conjugated polymers. The GIWAXS patterns for atactic PS and PMMA are depicted in **Figure 4**. Although the atactic PS and PMMA are commonly identified to be amorphous polymers, they still scatter diffusely in GIWAXS experiments and show some scattering peaks. However, these scattering peaks are all broad and isotropically orientated in all angles, which is reported to be the halo patterns of amorphous polymers[48], indicating the disordered amorphous or noncrystalline state and reflecting the near neighbor molecular distances that must be "discretized" locally and cannot be a continuous function on account of the molecular structure. Atactic PS shows two scattering peaks at $q$ ~0.71 Å$^{-1}$ and ~1.36 Å$^{-1}$, corresponding to an average spacing of 8.8 Å and 4.6 Å, respectively. After careful multi-peak fitting with Gaussian or Lorentzian distribution with a log cubic background, the fitted full width at half maximum (FWHM) of the two peaks are 0.48 Å$^{-1}$ and 0.5 Å$^{-1}$, respectively, corresponding to a $L_c$ of 13.1 Å and 12.6 Å. Comparing the $L_c$ values (13.1 Å and 12.6 Å) with nominal $d$-spacings (8.8 Å and 4.6 Å), the $L_c$ is only around two to three molecular layers thick, which is similar to that of a disordered liquid, indicating the disordered properties of PS. The $g$ parameter of the peaks at ~0.71 Å$^{-1}$ and ~1.36 Å$^{-1}$ of atactic PS is calculated to be 32.8% and 24.2%, respectively, indicating that atactic PS is indeed amorphous and even more amorphous than amorphous SiO$_x$. PMMA also shows a similarly broad, isotropically orientated scattering peaks that correspond to approximately two molecular layers for the peak located at ~ 0.96 Å$^{-1}$. The $g$ parameter is estimated to be 28%, again indicating an amorphous material. Meanwhile, the DSC of atactic PS and PMMA show no melting peak, but a pronounced glass transition, confirming that atactic PS and PMMA are 3D amorphous polymer with their properties summarized in **Table 1**. For conjugated polymers, there are some 3D amorphous polymers, such as regio-random P3HT and PTAA[19], with broad, featureless scattering peak in GIWAXS 2D patterns. With regards to nomenclature, 3D amorphous polymers are fully amorphous with the lack of ordering in any direction. The examples of PS and PMMA illustrate nicely that observation of a diffraction peaks does not indicate "ordering", much less any degree of "crystallinity", a fact worth keeping in mind when qualitative describing and analyzing GIWAXS data of semiconducting polymers.

For preferentially orientated ("2D") conjugated amorphous polymers, there is no observable melting peak in DSC. In addition, their GIWAXS patterns generally only exhibit a (100) lamellar peak and (010) π-π stacking peak, without any higher order ($h$00) peaks or (00$l$) backbone peaks. For example, the polymer FTAZ shows both the polymer alkyl stacking peak (100) at ~0.31 Å$^{-1}$ and π-π stacking peak (010) at ~1.65 Å$^{-1}$, as shown in **Figure 4**. The $L_c$ of (010) peak is 17.4 Å, which corresponds to around 4.6 molecular layers in the π-π stacking direction (details in **Table 1**), suggesting that the stacking in this direction is not extensive and only some local short-range order is present. The $g$ parameter related to the (010) peak is around 19%, implying low stacking order in this direction. Moreover, there is no melting peak observed in the DSC scan. Therefore, in spite of the local short-range order observed in the π-π stacking, there is no evidence for extensive ordering in other directions. Some other important and popular polymers such as PTB7-Th, PM6, PBDB-T (for details see **Table 1)** show similar GIWAXS and DSC characteristics, such as disordered (100) and short-range ordered (010) peak and with corresponding $g$ parameters >12%, an absence of the higher order diffraction peaks, and an absence of melting peak in DSC scans. We note that PBDB-T is somewhat an exception, in that it does exhibit higher order ($h$00) peaks, but the pseudo-Voigt mixing parameter of ($h$00) for PBDB-T is ~1 and the $g$ parameter is 19%. The strong disorder likely explains the lack of melting peak in DSC and makes it clear that the mere presence of higher order peaks is an insufficient qualitative characteristics to classify the molecular ordering. Accordingly, these polymers are also classified as preferentially oriented ("2D") amorphous polymers. We emphasize here that the ordering or packing is so limited in range (to 2-3 molecules) that we advocate to shun the use of "aggregate", let alone "crystallite", as the qualitative label to describe the spatial extent and arrangements of the packing.

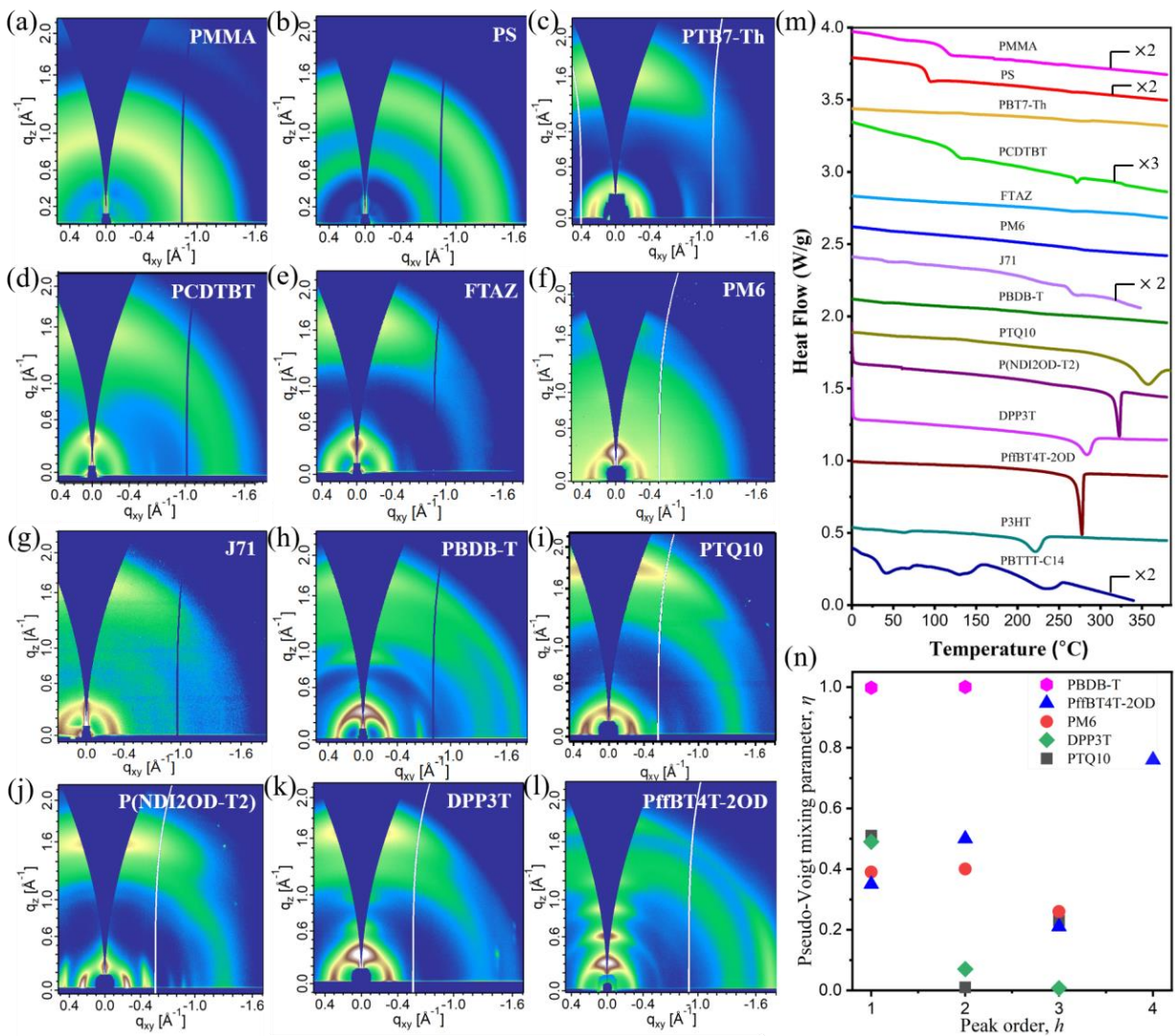

**Figure 4**. (a-l) The GIWAXS patterns of PMMA, PS, PTB7-Th, PCDTBT, FTAZ, PM6, J71, PBDB-T, PTQ10, P(NDI2OD-T2), DPP3T, PffBT4T-2OD, PBTTT, respectively. (m) The 1$^{st}$ heating of DSC thermograms with the heating rate of 10 °C/min of neat materials solution-cast from chlorobenzene (expect for PMMA, PS, PM6, PCDTBT, and PTQ10 which are cast from chloroform). The thermograms were vertically shifted and some profiles are enlarged to magnify the features. (n) Pseudo-Voigt mixing parameter, η, as a function of peak order of (h00).

For the semicrystalline conjugated polymers, we classify them into two types, the long-range ordered and short-range ordered. Both classes show a melting transition in DSC, although with different widths, and the GIWAXS yields different *g* parameters. A representative long-range ordered conjugated polymer is P3HT. Its GIWAXS 2D pattern shows distinctive (100), (200), (300) lamellar peaks observed in the out-of-plane direction with $q$ values of ca. 0.37 Å$^{-1}$, 0.77 Å$^{-1}$, 1.14 Å$^{-1}$, and well-defined π-π stacking peak (010) in the in-plane direction at $q$ ~1.68 Å$^{-1}$, corresponding a $d$-spacing of ~3.8 Å. The presence of the higher order ($h$00) peaks implies better molecular packing of the alkyl chains and better ordering properties than the previously discussed amorphous polymers that include FTAZ, and PTB7-Th. The $L_c$ of the P3HT (010) peak is 70.0 Å, which is around 18.4 molecular layers in the π-π stacking, exhibiting a relatively long-range order in this direction. The $g$ parameter for (010) peak is around 9%, indicating paracrystalline ordering. The DSC scan with a heating rate of 10 °C/min shows a pronounced melting peak at ~222.8 °C with a relatively large melting enthalpy of 19.7 J/g, suggesting the semicrystalline properties of P3HT. Based on these DSC and GIWAXS characteristics, *i.e.* long-ranged ordered π-π stacking, the presence of higher order ($h$00) peaks, and the observed melting signal in DSC, polymers such as PffBT4T-2OD and PBTTT are also characterized to be semicrystalline with long-range order. Details are provided in **Table 1**. Only in these cases would it be appropriate to use the qualitative label "crystallite" to describe the spatial arrangement and extent of the molecular packing.

Short-range ordered semicrystalline polymers also exhibit a DSC melting peak. In contrast to highly ordered polymers, there is a lack of long-range order as characterized with GIWAXS. There can be order along some crystallographic directions like alkyl stacking direction or backbone stacking direction, exhibiting higher order ($h$00) peaks or the presence of a narrow (00$l$) peak, but short-range semicrystalline polymers remain largely less ordered in at least one of the crystallographic direction, such as the π-stacking direction. In spite of the lower ordering in their π-stacks, they can form aggregates with local short-range order, providing effective pathways for intermolecular charge transport that results in much higher charge mobility than observed in amorphous polymer systems. A typical conjugated polymer with short-range order is DPP3T. The GIWAXS pattern of DPP3T shows discernable (100), (200), (300) lamellar peaks at $q$ values of around 0.33 Å$^{-1}$, 0.65 Å$^{-1}$, 0.95 Å$^{-1}$, respectively, and (010) π-π stacking peak in the out-of-plane direction at $q$ ~1.66 Å$^{-1}$, corresponding a $d$-spacing of ~3.7 Å. The $L_c$ of (010) peak is around 20.3 Å, which is approximately 5.3 molecular layers in the π-π stacking, implying a relatively short-range order in this direction. Moreover, the $g$ parameter for (010) peak is calculated to be 17%, suggesting amorphous properties. The DSC scan with a heating rate of 10 °C/min shows its melting point at ~284 °C with the melting enthalpy of 17.4 J/g. Therefore, although DPP3T shows semicrystalline character, its molecular packings and ordering extend only over a short-range. Another example is P(NDI2OD-T2) (also known as N2200), which not only shows higher order ($h$00) peaks, but also exhibits (001), (002), (003) backbone peaks in the GIWAXS 2D patterns, indicating good molecular packing. (NB: The absence of (00$l$) peaks in many other materials might not be due to disorder along the backbone, but a lack of scattering contrast between backbone moieties.) However, all peaks widths reflect less than 10 layers of packing in each direction, including in the π-stacking direction. Accordingly, P(NDI2OD-T2) is a short-range ordered semicrystalline polymer. We advocate that the most appropriate qualitative terminology to describe the spatial extent of the molecular packing is "aggregate".

PCDTBT is reported[40] to show the liquid crystalline properties, as we mentioned before. The GIWAXS of neat PCDTBT films shows only the first order of alkyl stacking peak (100) at ~0.35 Å$^{-1}$ and π-π stacking peak (010) at ~1.65 Å$^{-1}$ in the out-of-plane direction. The peak shape analysis performed at the (010) peak indicates that the $L_c$ of (010) peak is 17 Å, which is around 3.9 molecular layers of the π-π stacking, and the corresponding $g$ parameter is around 20% (see **Table 1**), suggesting that PCDTBT is amorphous. However, the DSC thermogram scan of PCDTBT shows a small melting peak at 271.5 °C with a very small enthalpy of 1.14 J/g, which is consistent with the reported nematic-isotropic transition temperature[40]. The small melting peak in DSC suggests that PCDTBT can be orientationally ordered although GIWAXS analysis would classify it as amorphous. It is thus the combination of DSC and WAXS that provides for complete classification.

**Table 1.** Summary of parameters characterized by DSC and GIWAXS for the conjugated polymers of high interest in organic electronics. Two amorphous insulting polymers PS and PMMA were also evaluated for comparison.

| Polymer | Melting temperature $T_m$ ($\Delta T^a$) [°C] | Melting enthalpy $\Delta H_m$ [J/g] | order of lamellar packing ($h$) | π-π stacking distance ($d$) [Å] | Coherence length of π-stacking ($L_c$) [Å] | Number of π-π stacked layers ($n$) | $g$ for (100) peak$^d$ | $g$ for (010) peak$^d$ | Remark |
|---|---|---|---|---|---|---|---|---|---|
| PMMA | N/A | N/A | 0 | 6.6$^b$ | 13.7 | 2.1 | N/A | 28%$^b$ | 3D amorphous |
| PS | N/A | N/A | 0 | 4.6$^b$ | 12.6 | 2.7 | N/A | 24%$^b$ | 3D amorphous |
| PTB7-Th | N/A | N/A | 1 | 3.9 | 15.3 | 3.9 | 28% | 20% | Oriented ("2D") amorphous |

| | | | | | | | | |
|---|---|---|---|---|---|---|---|---|
| PCDTBT | 271.5 (10.5) | 1.14 | 1 | 4.4[49] | 17[49] | 3.9 | 32%[49] | 20% | Oriented ("2D") amorphous |
| FTAZ | N/A | N/A | 1 | 3.8 | 17.4 | 4.6 | 23% | 19% | Oriented ("2D") amorphous |
| PM6 | N/A | N/A | 3 | 3.7 | 18.2 | 4.9 | - | 18% | Oriented ("2D") amorphous |
| PBDBT | N/A | N/A | 2 | 3.7 | 18.8 | 5.1 | 19% | 18% | Oriented ("2D") amorphous |
| P(NDI2OD-T2) | 312.3 (12.0) | 15.9 | 4 | 3.9 | 15.5 | 4.0 | 4%[13] | 17% | Short-range ordered (aggregate) |
| DPP3T | 283.5 (27.1) | 17.4 | 3 | 3.8 | 20.3 | 5.3 | - | 17% | Short-range ordered (aggregate) |
| PTQ10 | 356.7 (47.3) | 17.7 | 3 | 3.5 | 33.1 | 9.5 | - | 13% | Short-range ordered (aggregate) |
| PffBT4T-2OD | 277.8 (8.9) | 21.5 | 4 | 3.5 | 48.3 | 13.8 | - | 11% | Long-range ordered (crystallite) |
| P3HT | 222.8 (18.3) | 19.7 | 3[50] | 3.8[50] | 70.0[50] | 18.4 | 8%[19] | 9% | Long-range ordered (crystallite) |
| PBTTT | 230.8 (60.6) | 23.2 | 4[51] | 3.7[51] | 82.6[51] | 22.3 | 7%[51] | 8% | Long-range ordered (crystallite) |
| TIPS-Pentacene[c] | 261 (4.3)[52] | 25.6[52] | 4[13] | 7.8[53] | 897.1[53] | 115.0 | 0.3%[13] | 4% | Long-range ordered (crystallite) |

[a]$\Delta T$ refers to the peak width of the melting peak. [b]Since there is no π-π stacking peak for PS and PMMA, the peak used here is their amorphous peak with the highest intensity. [c]The data of the 4$^{th}$ and 8$^{th}$ columns for TIPS-Pentacene are shown for (00$l$) stacking. [d]Here, all $g$ parameters are calculated with single-peak width analysis, which is a good estimation of the paracrystallinity when the effect from lattice-parameter fluctuation $e_{rms}$ can be neglected[13].

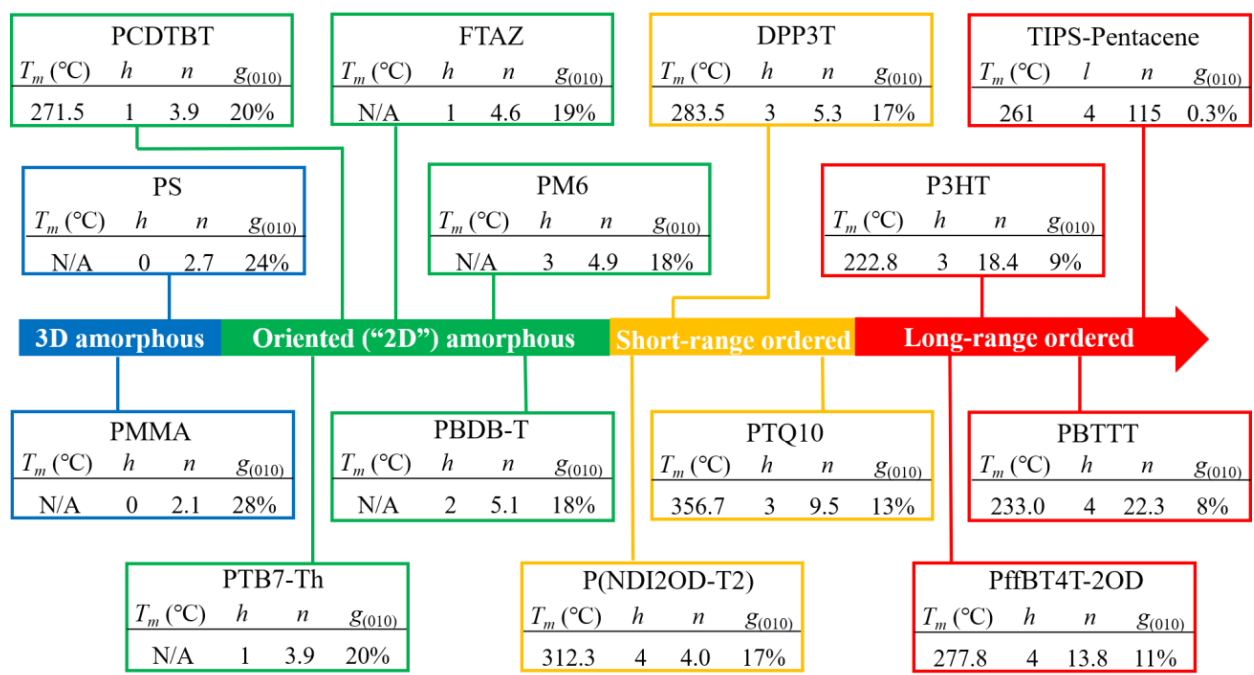

**Figure 5**. Summary of over 10 representative polymers on a disorder-order scale characterized by the *g* parameter. Relevant parameters such as $T_m$, *h* (*l*), *n*, and $g_{(010)}$ refer to the melting temperature (in Celsius), the presence of the highest order of lamellar (backbone) stacking, the number of π-π stacked layers, and the paracrystalline disorder parameter for (010) peak, respectively.

## 4. Toward Polymer Crystals with *g* <5% and Long Coherence Length

The prior section has made it clear that even the most highly ordered polymers to date have a *g* parameter of >6% except for the TIPS-pentacene with a super low *g* parameter and well-defined triclinic unit cell, and the question that begs is simple: Can semiconducting polymeric materials with significantly lower *g* parameters be designed? In this section, we are going to discuss the challenging task of pursuing perfect semiconducting polymer crystals with low lattice disorder and possible approaches to achieving the goal.

### 4.1 Why it is essential to pursue perfect polymer crystals

The molecular order of semiconducting polymers has been extensively studied and correlated with the charge transport properties, such as the exciton diffusion length and charge mobility, which are essential for high-performance devices. It is reported that the exciton diffusion length is a monotonic function of the extent of molecular order and its impact on energetic disorder[54]. X. Jin et al.[55] recently reported that the exciton diffusion length can be as high as 200 nm in conjugated polymer nanofibers prepared by seeded growth to form well-ordered domains with increased molecular ordering and low energetic disorder. On the other hand, the charge mobility and material conductivity, are also reported to be related to the molecular order, such as the π-π stacking distance[56], the coherence length[57, 58], and the degree of crystallinity[56]. The charge mobility can vary by several orders of magnitude and the highest charge mobility can be found in single crystals[59]. Some commonly recognized nearly amorphous polymers might also show good charge transport if they exhibit stacked aggregates with good connectivity, allowing locally efficient intermolecular charge transport and providing sufficient pathways for charge transport[19]. Forming a perfect crystal is an essential way to improve the charge transport properties of these polymers.

However, the crystallites formed in the polymeric materials are usually far away from the perfect crystals, lacking ordering in one or two crystallographic axes, limiting the size of the crystallites and thus the coherence length of the stackings, especially of the π-stacking since the increase in $L_c$ and DoC along the lamellar stacking direction has little impact on the electron transport properties due to the insulating nature of the solubilizing side chains[60]. The coherence length of the π-π stacking is generally reported to be tens of intermolecular spacings. Interestingly, J.H. Carpenter et al.[61] reported recently that the lamellar packing of FTAZ shows an extremely long range of order within the highly ordered side-chain layer with a coherence length > 70 nm (limited by instrument broadening) that corresponds to more than 150 intermolecular spacings. This opens up an opportunity and possibility to pursue perfect crystals with long range molecular order in all directions, including backbone and π-stacking, if this ordering can be achieved synergistically for the backbone.

### 4.2 Challenges of perfect crystals

A perfect crystal of semiconducting conjugated polymers is expected to show excellent ordering along all three crystallographic axes, which are the π-π stacking direction, side-chain lamellar packing direction, and the backbone direction. However, the ordering in the backbone and π-stacking directions are not necessarily achieved synergistically with ordering in the lamellar stacking direction or ordering between alkyl side chains. There exist different types of ordering configuration away from the perfect crystals, discussed in sequence below and shown in **Figure 6**.

First, the polymer chains are well oriented but less ordered. A representative case is the liquid crystalline polymers, such as PBTTT where the in-plane PBTTT crystal orientation varies smoothly across a length scale, significant in a relatively long range while only small angle variations between adjacent diffracting regions, exhibiting an in-plane liquid crystalline phase[62]. Second, there are often cases when the orientation of the π-stacking is comprising of both edge-on and face-on to the substrate, which is also called bimodal texture[63], impeding the optimization of the molecular order. H. Kim *et al.*[64] reported that films with predominantly edge-on texture show much improved crystalline ordering than the bimodal texture. This backbone orientation can be altered by the tuning in the side chain. Moreover, there are generally a large number of grain boundaries, which tend to limit the size of the ordered phase and impede the charge transport.

Last but not least, there can be a competition between backbone ordering and sidechain ordering that results either in an ordered sidechain layers with a disordered backbone layers or ordered backbones with disordered sidechains. J.H. Carpenter *et al.*[61] has indeed reported this competition in a variety of semiconducting conjugated polymers. FTAZ, which we classified above as a 2D amorphous polymer with $g_{(010)}$=19% can, for example, under certain processing conditions exhibit unusually long $L_c$ for lamellar ordering that induces significant torsional backbone disorder, resulting in a vertically multilayered composite nanostructure consisting of the ordered sidechain layers alternating with disordered backbone layers or vice versa. The sidechain ordering in FTAZ had characteristic GIWAXS, DSC, optical and near edge x-ray absorption fine structure (NEXAFS) signature. Upon melting of the highly ordered sidechain phase, the backbones are no longer torsionally constrained and thus able to π-stack more efficiently, yielding better edge-on orientated π-stacking and improved charge carrier mobility. NEXAFS spectroscopy could demonstrate this competing ordering, as shown in **Figure 7**. The NEAXFS spectra of drop-cast FTAZ (**Figure 7a**) show two peaks at 287.4 and 288.1 eV, corresponding to C 1s → $\sigma^*_{C-H}$ transitions that agree well with the published NEXAFS for a nonadecane crystal[65], demonstrating the high degree of sidechain ordering further corroborated by strong angular dependence of the NEXAFS intensity. Considering the C 1s → $\pi^*_{C=C}$ transition region from ~ 283.5-287 eV related to the backbone, only a single, broad peak is observed with limited angular dependence. This suggests a disordered backbone with near random orientation distribution. In contrast, the NEAXFS spectra of spin-cast FTAZ (**Figure 7b**) show the opposite characteristics with highly ordered backbone (splitting peak in the C 1s → π*$_{C=C}$ transition) but disordered sidechain with a single peak for the C 1s → σ*$_{C-H}$ transitions that do not exhibit any angular dependence. This competition between the sidechain and the backbone ordering is also observed in PCDTBT, N2200 and P3HT films and might therefore be a general feature of most semiconducting materials to date.

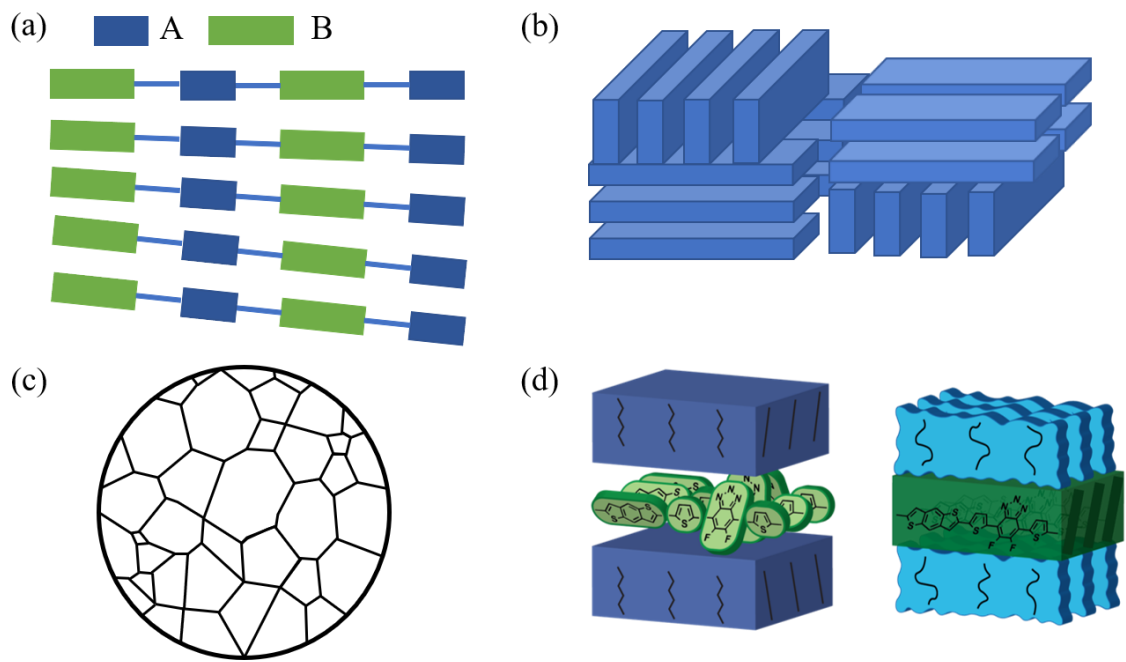

**Figure 6**. Schematic of the different types of molecular packing and textures that depart from the perfect single crystals. (a) well oriented but less ordered packing (liquid crystalline ordering), the polymer chains are well oriented but the adjacent chains are rotating gradually, impacting the molecular stacking; (b) There are both edge-on and face-on orientation of the packing. (c) There exists the grain boundary between two well-ordered domains. (d) The competition between the backbone ordering and the side-chain ordering. Reproduced with permission from ref. 61.

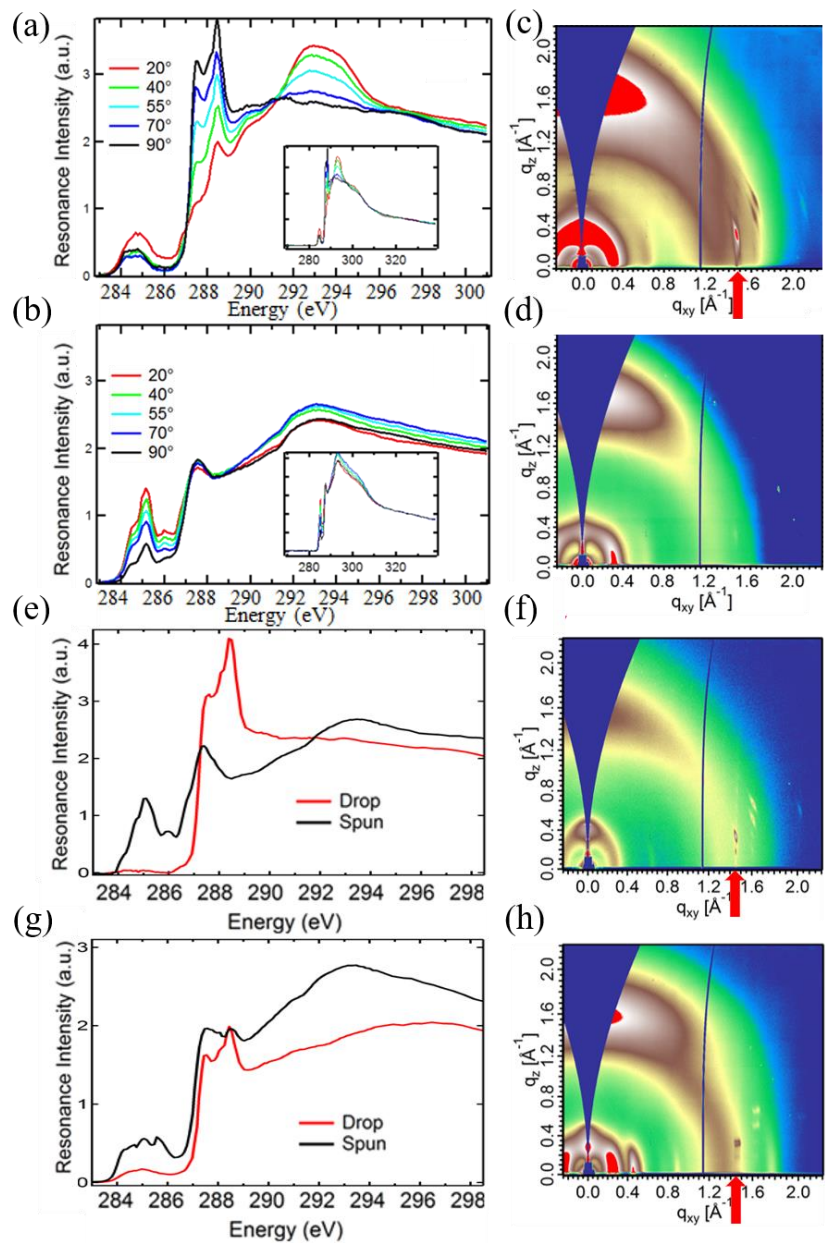

**Figure 7.** NEXAFS spectra that demonstrates the competition between backbone ordering and sidechain ordering and GIWAXS patterns that show the high degree of sidechain ordering for drop-cast thin films from chlorobenzene. Reproduced with permission from ref. 61. The NEXAFS spectra for (a) drop-cast FTAZ and (b) spin-cast FTAZ, where 90° is normal incidence. GIWAXS patterns for (c) drop-cast FTAZ and (d) spin-cast FTAZ. NEXAFS spectra at normal incidence of drop-cast and spin-cast films of (e) PCDTBT, and (g) N2200. 2D GIWAXS patterns from drop-cast films of (f) PCDTBT, and (h) N2200. The sharp features at ~ 1.47 Å$^{-1}$ highlighted with red arrows in drop-cast films indicate the high degree of sidechain ordering.

## 4.3 Possible solutions towards single crystals with low *g* parameters

Although there are a lot of challenges to pursue the perfect crystals, there are some possible solutions. Two aspects need to be separately considered: the size of the crystals and degree of crystallinity one can achieve, and the lattice disorder as captured in the *g* parameter. Solvent engineering is a facile approach to induce large crystals for semicrystalline polymers. For instance, C. Müller *et al.*[66] induced macroscopic-sized and highly ordered crystalline domains of P3HT (**Figure 8a**) via a mixture of crystallizable solvent 1,3,5-trichlorobenzene (TCB) and a second carrier solvent such as chlorobenzene. The solidification was initiated by growth of macroscopic TCB spherulites and followed by replicated epitaxial crystallization of a variety of conjugated polymers, such as P3HT and PCPDTBT, on TCB crystals. When TCB is removed, the macroscopic-sized P3HT or PCPDTBT spherulites were left behind. Expect for the enlarged size of polymer crystal, the induced spherulites also show more preferentially face-on orientation and decreased *g* parameter for the π-π stacking as estimated here from the narrower diffraction peaks. Another way to improve the molecular order is the strain-induced alignment. L. H. Jimison *et al.*[67] reported that P3HT can be directionally crystallized in micrometers (**Figure 8b**) with a lower *g* parameter with the help of TCB which at first acts as a solvent and then as a substrate for polymer epitaxy after TCB solidifies in characteristic needle-like crystals. X. Jin *et al.*[55] suggested to incorporate the nanoparticles into the polymer nanofibers (**Figure 8c**) to improve exciton diffusion and charge transport, which is correlated with the improved ordering for the conjugated polymer confirmed by WAXS data and the pronounced vibronic structure and narrow spectral linewidths in the photoluminescence. B. Kang *et al.*[68] demonstrated that the strong self-organization of the semifluoroalkyl side chains in PNDIF-T2 and PNDIF-TVT induced the identical unit-cell structure of the polymer crystallites despite their different backbone moieties with different size, resulting in the long-range ordering and low *g* parameters of 10.6% and 9.0%, respectively, The x-ray scattering data showing the tight interdigitation of antiparallel semifluoroalkyl chains and rigid polymer backbone suggests a superstructure consisting of conjugated backbone crystals within a crystal of fluoroalkanes. This coexistence of the backbone crystallites and the sidechain crystallites suggests a possible way to reach the perfect crystals. J.H. Carpenter *et al.*[61] also speculated that stereochemistry control is the key for perfect ordering for high performance to explore the synergistic ordering of the backbone and side chain with the aid of sufficient molecular modeling and design. The reported exceptionally long-range sidechain ordering could be utilized to facilitate similar long-range backbone ordering. This stereochemistry control is also favored by D. Venkateshvaran *et al.*[69] to approach the "disorder free" self-assembly since they reported that the low degree of energetic disorder of IDTBT originates in the remarkable resilience of the torsion-free backbone conformation to the side-chain disorder.

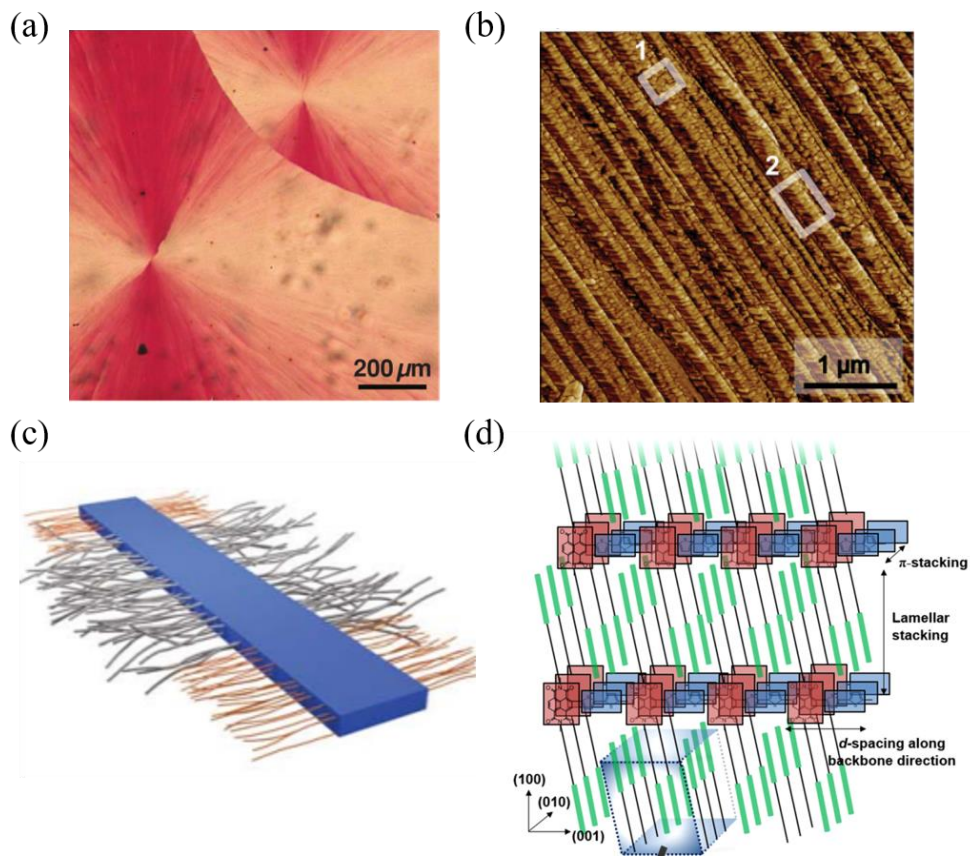

**Figure 8**. (a) The spherulite-like structure of P3HT:PC$_{61}$BM formed as a replica of the original TCB spherulites. Reproduced with permission from ref. 66. (b) The micrometer-sized directionally crystallized P3HT induced by mechanical strain. Reproduced with permission from ref. 67. (c) Schematic of the highly ordered nanofibers formed with the seeded growth process. Reproduced with permission from ref. 55. (d) The superstructure of "backbone crystals" and "side-chain crystals" via side-chain-induced self-organization. Reproduced with permission from ref. 68.

## 4.4 Conceptual backbone graft hetero-epitaxy onto side-chain crystals: A new molecular design paradigm

We note that only a few materials have been able to be made to crystalize to date, all of which still have $g$ parameters > 6% in at least one direction. We deduce from the preponderance of the literature and the discussion above that this is due to the intrinsic molecular design, and specifically due to the current incommensurability of the ideal sidechain spacing with the spacing determined by the available attachment points along the backbone. In other words, the spacing and density of the attachment points of the sidechains for a layer of perfectly ordered backbones do rarely corresponds to the spacing and density in a fully ordered sidechain layer. PBTTT is a rare exception that comes close to "commensurability". Taking PBTTT-C$_{14}$ as an example, it is reported[70] for alkane crystal of C$_{14}$H$_{30}$ to show a triclinic structure with unit-cell parameters a = 4.29 Å, b = 4.82 Å, c = 19.84 Å, α = 84.10°, β = 66.82°, γ = 73.00°, while the distance of the sidechain attachment point in the backbone is approximately 9.5 Å, which is around twice of b parameter of the alkane crystal of the sidechain but slightly less. This would imply that for perfect side chain ordering, the PBTTT would be slightly in compression, which likely causes some backbone disorder. Importantly, the results of J.H. Carpenter *et al.*[61] have shown that the side-chain ordering dominates energetically over the backbone ordering at room temperature even in cases of nominally amorphous polymers such as FTAZ where the distortions of the backbone must be energetically costly. We suggest that this strong ordering of the sidechains should be exploited and that the design and synthesis of semiconducting polymers borrow the concepts of heteroepitaxy from compound semiconductor thin film growth. It is likely instructive and advantageous to reverse the design considerations. While it is the backbone design that is currently contemplated first prior to synthesis and then various sidechains are attached wherever synthetically convenient, the reverse might be interesting if not tantalizing to consider.

Let us create a conceptual perfect 2D crystal layer of sidechain *in-silico* of which spacing can be slightly varied by the length and nature of the sidechains (alkene, ethylene glycol, floriated alkanes, etc.). Subsequently, we design a backbone and possible linkers *in-silico* to conceptually graft said backbone commensurably with this perfect alkane layer in such a way that the backbone is also perfectly ordered. A material designed in this way should be much more readily ordered during processing into 2D nano-composites of conducting and insulating layers. Most excitingly, the stronger tendency for the sidechain rather than backbones to order at RT might be used to strain the backbone or to control the stacking and thus the transfer integral between polymer chains. This paradigm might allow totally new ways of designing electronic properties in the 2D layers that are separated by an insulating layer. We note here that use of any branched sidechains without control of stereochemistry would likely destroy the ability to achieve "hetero-epitaxy".

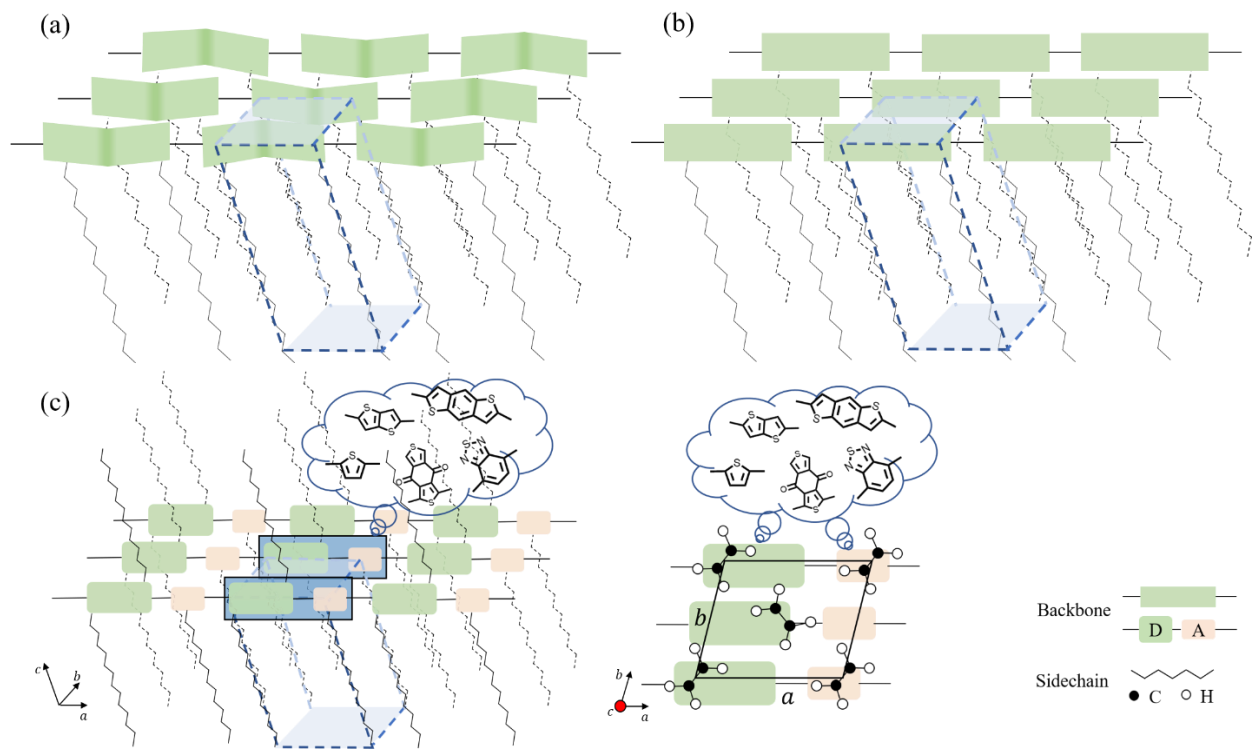

**Figure 9.** Schematic illustration of the graft hetero-epitaxy concept for the chemical structure design for conjugate polymers. With the unit-cell parameter of a perfect alkane crystal of sidechains, the sidechain attachment points at the backbone are determined. (a) When the size and density of attachment of backbone moiety is not commensurate with the lattice parameter of sidechain crystals, the backbone is compressed or stretched, and thus the backbone ordering is reduced. (b) If there is commensurability between the backbone and sidechain packing, the backbone and sidechain ordering can be reached synergistically, making it possible to reach perfect conjugate polymer crystals. To demonstrate how to design this commensurability, (c) shows the design strategy that the backbone moieties need to be carefully chosen or designed to match with the size and density of attachment points, dihedral angle, etc. of the alkane crystal of sidechains.

**Summary and Outlook**

In summary, we have clarified qualitative nomenclature related to molecular order that are commonly used in the organic electronics, such as crystalline, semicrystalline, amorphous, crystallinity, crystallites, and aggregates, and advocated a specific usage. With the help of GIWAXS and DSC, one can unambiguously identify the crystalline properties of semiconducting polymers ranging from 3D amorphous to long-range ordered paracrystalline polymers with crystallites. We specifically advocate that values for *g* parameters are broadly reported to allow for a more consistent use of terminology and comparison between materials and the literature at large. We have also contemplated how to improve the molecular ordering of the semiconducting conjugated polymers to approach perfect crystals as an essential way to improve the charge transport properties and thus the device performance. Although there are lots of unsolved problems, further advances in semiconducting conjugated polymer design and synthesis might lie in the proposed hetero-epitaxy concept and stereochemistry control to explore the synergistic ordering of both the backbone and the sidechains in thermodynamically favorable packing configurations or tune the side chain crystallinity and interdigitation to provide a mechanism for three-dimensional crystallization, improving the ordering and thus yielding higher performance.

**Glossary**

*Aggregate*: The aggregate in solution or thin films will exhibit the presence of red-shifted or blue shifted absorption peak in the absorption spectrum compared to the unaggregated state.

*Amorphous*: Amorphous materials are characterized by the absence of molecular ordering and any melting peak in DSC. Here, we classify the amorphous materials into two type: 3D amorphous and oriented ("2D") amorphous. For 3D amorphous materials, it is scattered diffusely in X-ray diffraction experiments with broad and isotopic scattering peaks. For oriented ("2D") amorphous materials, there is no extensive ordering in other directions except for the local short-range order observed in the π-π stacking direction.

*Backbone ordering*: The ordering of polymer backbones on a well-defined lattice.

*Bragg reflection*: The reflection that satisfies Bragg condition, $2d\sin\theta = n\lambda$, where d is the stacking distance, θ is the incidence angle, λ is the wavelength of the X-ray and n is an integer called the order of diffraction.

*Coherence length ($L_c$)*: It is the length scale that related to peak width via Scherrer equation: $L_c=2\pi K/\Delta q$, where K is a shape factor (typically 0.8-1) and $\Delta q$ is the FWHM of a diffraction peak.

*Conformational freedom*: The spatial or structural arrangement of atoms affording distinction between stereoisomers can be interconverted by rotations about formally single bonds.

*Configurational freedom*: The spatial relation of atoms in the molecule is not fixed, e.g. different chirality or tacticity.

*Crystalline*: Crystalline state is characterized by three-dimensional, long-range order on an atomic scale.

*Crystallite*: The ordered domains with small size on the order of 1-100 nm.

*Crystallinity*: degree of crystallinity (DoC), the volume fraction of crystalline domains in semicrystalline films.

*Cumulative disorder*: Statistically homogeneous disturbances to an ideal crystalline lattice which produce long-range distortion, including dislocations, impurities, chain backbone twists, or nonideal packing.

*Debye-Waller factor*: Describing the attenuation of x-ray scattering or coherent neutron scattering caused by thermal fluctuation.

*Defect*: The defect in a crystal includes point defects, linear defects and planar defects. Point defects refer to places where an atom is missing or irregularly placed in the lattice structure, including vacancies and impurity. Linear defects (or dislocation) are groups of atoms in irregular positions. Planar defects are interfaces between homogeneous regions of the material, such as grain boundary.

*TEM*: transmission electron microscopy, a microscopy technique where a beam of electrons is transmitted through a specimen to form an image. The interactions between the electrons and the atoms can be used to observe nanoscale features such as the crystal structure and features in the structure like dislocations and grain boundaries.

*Diffraction peak*: the diffraction pattern acquired from the coherently scattered intensity.

*DSC*: differential scanning calorimetry, used to detect the difference between the heat flows of a specimen cell and a reference cell that are simultaneously exposed to the same heating flux, detecting the amount of heat absorbed or released from the material during the heating or cooling to record the thermal transition like the glass transition or melting of polymer crystals.

***Enthalpy of fusion***: the change in the enthalpy resulting from providing energy, typically heat, to a specific quantity of the substance to change its state from a solid to a liquid, at constant pressure.

***Ewald sphere***: Ewald's sphere is drawn with its center at the origin of the $\mathbf{k}_o/(2\pi)$ vector and radius of $1/\lambda$ where $\mathbf{k}_o$ is the incoming wave vector in a given reciprocal lattice and $\lambda$ is the wavelength. Any reciprocal lattice point that lies on this sphere can be connected by reciprocal lattice vector $\mathbf{G}_{hkl}$, satisfying Bragg condition.

***FWHM***: full width at half maximum.

***Gaussian distribution***: normal distribution, a type of continuous probability distribution for a real-valued random variable. The Fourier transform of a Gaussian function is a Gaussian.

***GIWAXS***: grazing-incidence wide-angle X-ray scattering, a structural measurement technique wherein wide-angle scattering and molecular length scales are collected.

***g parameter***: paracrystalline disorder parameter, a measure of the percentage of statistical deviation from the mean lattice spacing in a crystal.

***H-aggregate***: The main absorption peak is blue-shifted due to the side-by-side orientation between neighboring stackings.

***Hetero-epitaxy grafting***: the growth of a crystalline film based a well-defined orientation with respect to the crystalline substrate of a different material. Here, we borrow this concept to refer to a proposed conjugated polymer design strategy that the backbone moieties are chosen or design to be commensurate with the lattice parameters of sidechain crystals.

***J-aggregate***: Characterized by the appearance of an intense, red-shifted narrow band in the absorption spectrum due to head-to-tail orientation between neighboring stackings.

***Lattice-parameter fluctuations ($e_{rms}$)***: the variance of the interplanar spacing within a sample (from one crystallite to the next, from one area of the diffracting volume to another), characterizing inhomogeneities within a sample, such as a slight contraction or expansion of the lattice spacing due to appearance of interfaces.

***Line shape analysis***: an analysis based on trends of peak widths and Lorentzian components of pseudo-Voigt line shapes as a function of diffraction orders. The peak width as a function of peak order can help distinguish the contribution of finite size from the cumulative disorder to the broadening of peaks while the Lorentzian components of pseudo-Voigt line shapes is a good way to determine the paracrystalline disorder from lattice parameter fluctuation.

***Liquid crystalline***: Liquid crystalline materials show the properties of both a crystal and a liquid, which is in an intermediate state between the amorphous state and crystalline state.

***Lorentzian distribution***: The Fourier transform of an exponentially decaying function.

***Morphology:*** Characterizing the nanostructure or microstructure of a bi-continuous network of domains in terms of molecular packing and phase separation.

***NMR***: nuclear magnetic resonance. When nuclei in a strong constant magnetic field are perturbed by a weak oscillating magnetic field, an energy transfer is possible between the base energy to an excited energy level and then emit an electromagnetic signal with the same frequency when it returns to the base energy level. NMR is sensitive to local chemical environment and thus used to probe local interaction and ordering.

***Noncumulative disorder***: Random statistical fluctuations about an ideal lattice position.

***Number of stacking layers (n)***: Characterized by the ratio of the coherence length and the stacking distance of a certain diffraction peak.

***Packing***: The arrangement of molecular chains.

***Paracrystalline***: Paracrystalline state is characterized by lattice distortions or uncorrelated displacements of the atoms away from their average lattice spacing and hence by limited order along given lattice directions. The paracrystalline disorder is quantified by paracrystallinity parameter $g$.

***Pole figure***: A pole figure is a plot of the orientation distribution of a particular set of crystallographic reciprocal lattice planes, providing a useful illustration of a material's texture.

***Pseudo-Voigt mixing parameter $\eta$***: The fraction of Lorentzian function in a Voigt profile. $\eta = 0$ represents a Gaussian line shape, and $\eta = 1$ represents a Lorentzian.

***P-SoXS***: soft x-ray scattering with polarized light. P-SoXS is sensitive to bond orientation which lies on the orientational material contrast between the cases where the average dipole moment is aligned parallel or perpendicular to the incident electric field.

***Scattering peak***: The incoherent scattering signal due to the Compton scattering.

***Semicrystalline***: Semicrystalline materials are composed of ordered domains but also a volume fraction of amorphous domains.

***Sidechain ordering***: The ordering of polymer sidechains.

***Specific volume***: The number of cubic meters occupied by one kilogram of matter, characterized by the ratio of a material's volume to its mass.

*Stacking distance (d)*: Characterized by the reciprocal of the peak position (q) with $d=2\pi/q$. It describes how close the molecules pack together.

*Texture*: The crystallite alignment or orientation on the order of 1-100 nm.

*Warren-Averbach framework*: It employs the deconvolution Fourier-transform method for the determination of the intrinsic physical line profile, followed by the Fourier method for evaluation of lattice imperfections. This method states that the numerically calculated Fourier coefficients for intrinsic physical line profile are the product of two terms: the size contribution (independent of peak order) and the disorder coefficient that are peak-order dependent.

**Appendix A: the full name of polymers used in the manuscript**

**DPP3T**: Poly{2,2′-[(2,5-bis(2-hexyldecyl)-3,6-dioxo-2,3,5,6- tetrahydropyrrolo[3,4-c ]pyrrole-1,4-diyl)dithiophene]- 5,5′-diyl-alt-thiophen-2,5-diyl}

**FTAZ**: Poly[(3-butylnonyl)benzodithiophene-fluorinatedtriazole]

**J71**: Poly[[5,6-difluoro-2-(2-hexyldecyl)-2H-benzotriazole-4,7-diyl]-2,5-thiophenediyl[4,8-bis[5-(tripropylsilyl)-2-thienyl]benzo[1,2-b:4,5-b']dithiophene-2,6-diyl]-2,5-thiophenediyl]

**P3HT**: Poly(3-hexylthiophene-2,5-diyl)

**PBDB-T**: Poly[(2,6-(4,8-bis(5-(2-ethylhexyl)thiophen-2-yl)-benzo[1,2-b:4,5-b']dithiophene))-alt-(5,5-(1',3'-di-2-thienyl-5',7'-bis(2-ethylhexyl)benzo[1',2'-c:4',5'-c']dithiophene-4,8-dione)]

**PBDT-TDZ**: Poly[1,3,4-thiadiazole-(benzo[1,2-b:4,5-b′]dithiophene)]

**PBDTS-TDZ**: Poly[1,3,4-thiadiazole-2,5-diyl(3-octyl-2,5-thiophenediyl)[4,8-bis[(2-butyloctyl)thio]benzo[1,2-b:4, 5-b′]dithiophene-2,6-diyl](4-octyl-2,5-thiophenediyl)]

**PBTTT**: Poly[2,5-bis(3-tetradecylthiophen-2-yl)thieno[3,2-b]thiophene]

**PCDTBT**: Poly[N-9'-heptadecanyl-2,7-carbazole-alt-5,5-(4',7'-di-2-thienyl-2',1',3'-benzothiadiazole)]

**PffBT4T-2OD**: Poly[(5,6-difluoro-2,1,3-benzothiadiazol-4,7-diyl)-alt-(3,3'''-di(2-octyldodecyl)-2,2',5',2'',5'',2'''-quaterthiophen-5,5''''-diyl)]

**PM6**: Poly[(2,6-(4,8-bis(5-(2-ethylhexyl-3-fluoro)thiophen-2-yl)-benzo][1,2-b:4,5 -b'] dithiophene))-alt-(5,5-(1',3'-di-2-thienyl-5',7'-bis(2-ethylhexyl)benzo[1',2'-c:4',5' -c']dithiophene-4,8-dione)

**PMMA**: Poly(methyl methacrylate)

**P(NDI2OD-T2)**: Poly{[N,N'-bis(2-octyldodecyl)naphthalene-1,4,5,8-bis(dicarboximide)-2,6-diyl]-alt-5,5'-(2,2'-bithiophene)}

**PS:** Polystyrene

**PTAA**: Poly(triaryl amine)

**PTB7**: Polythieno[3,4-b]-thiophene-co-benzodithiophene

**PTB7-Th**: Poly[4,8-bis(5-(2-ethylhexyl)thiophen-2-yl)benzo[1,2-b;4,5-b']dithiophene-2,6-diyl-alt-(4-(2-ethylhexyl)-3-fluorothieno[3,4-b]thiophene-)-2-carboxylate-2-6-diyl)]

**PTQ10**: Poly[[6,7-difluoro[(2-hexyldecyl)oxy]-5,8-quinoxalinediyl]-2,5-thiophenediyl]

**TIPS-Pentacene**: 6,13-Bis(triisopropylsilylethynyl)pentacene

## Acknowledgements

The authors would like to acknowledge the support from the ONR grant N000141712204. GIWAXS data were acquired at beamline 7.3.3 at the Advanced Light Source, which are supported by the Director of the Office of Science, Office of Basic Energy Sciences, of the U.S. Department of Energy under Contract No. DE-AC02-05CH11231. The authors gratefully acknowledge Subhrangsu Mukherjee, Masoud Ghasemi, Indunil Angunawela for the acquisition of part of GIWAXS and DSC data and appreciate the fruitful discussions with Reece Henry.